\documentclass[aps,prb,twocolumn,psfig,groupedaddress,floats,showpacs]{revtex4}
\usepackage{amsmath}
\usepackage{amssymb}
\usepackage{graphicx}
\usepackage{dcolumn}
\usepackage{natbib}
\usepackage{bm}
\usepackage{epsfig}
\usepackage{amsfonts}
\usepackage{color}
\usepackage{hyperref}%
\providecommand{\U}[1]{\protect\rule{.1in}{.1in}}
\providecommand{\U}[1]{\protect\rule{.1in}{.1in}}
\providecommand{\U}[1]{\protect\rule{.1in}{.1in}}
\providecommand{\U}[1]{\protect\rule{.1in}{.1in}}
\providecommand{\U}[1]{\protect\rule{.1in}{.1in}}

\begin{document}
\title{A Quantum Spin Hall Round Disk as a Spin Rotator and Filter}
\author{Zhan-Feng Jiang and Wen-Yu Shan}
\affiliation{Department of Physics, and Center of Theoretical and Computational Physics,
The University of Hong Kong, Pokfulam Road, Hong Kong}
\date{\today}

\begin{abstract}
We study theoretically the spin transport of a Quantum Spin Hall round disk.
When an electron traverses the disk in virtue of the edge states, its spin's
in-plane component can be rotated by a magnetic flux through the disk. The
spin rotation occurs due to the interference of two helical edge states with
opposite spins, which is regarded as the Aharonov-Bohm effect in the spin
space and a manifestation of the Berry phase. Besides, the disk has a spin
filter effect on the tunneling current when we apply an appropriate magnetic
field and gate voltage on it. The spin polarization ratio can reach 100\% when
the couplings between the disk and leads are weak.

\end{abstract}

\maketitle

\section{Introduction}

The geometric phase of quantum system undergoing adiabatic, cyclic evolution
was first discovered by Berry \cite{Berry} and a simple geometric
interpretation was given by Simon \cite{Simon}. It has a tremendous impact on
various areas of physics which have triggered active researches over a quarter
of century. Now it is recognized that the geometric phase is an important
concept of quantum mechanism both from a theoretical perspective and potential
applications. A manifestation of the Berry phase is the well-known
Aharonov-Bohm (AB) effect \cite{AB} that an electrical charge which cycles
around a magnetic flux. From the viewpoint of the symmetry between magnetic
field and electric field in the Maxwell equations, Aharonov and Casher
predicted that a magnetic moment acquires a phase around a charge flux line
\cite{AC}. The the Aharonov-Casher (AC) effect is weak in the neutron
interference experiment \cite{AC exp}, while recently an enhanced AC effect is
realized in the two-dimensional electron system confined in a semiconductor
quantum well with strong spin-orbit coupling \cite{Nitta,Sinova exp}. This is
another great manifestation of the Berry phase.

The researches on the spin-orbit interaction system \cite{Zhangshoucheng,
Niuqian} lead to the discovery of the quantum spin Hall (QSH) material
\cite{QSH,experiment} as a new phase of the condensed matter, which can be
regarded as two copies of the integer quantum Hall systems for up and down
spins with the opposite chiralities. For a strip geometry, the edge states for
up and down spins propagate in the opposite directions along each edge, which
are called the helical edge modes \cite{N1,N2,N3,N4}. When the Fermi energy of
a QSH system is in the bulk gap, only the edge modes are responsible for
transport, thus the whole system can be regarded as two one-dimensional (1D)
channels on both edges of the strip.

Recently, Chu \textit{et.al.} has predicted the AB oscillations in the
magnetoconductance of a singly connected QSH disk \cite{Churuilin}. In this
paper, we consider the spin transport of a QSH round disk connected to two
normal electrodes with a vanishing spin-orbit interaction (SOI). The system is
equivalent to a 1D AB ring in which the transport is carried out by the
helical edge states. The up and down spin propagating in opposite directions
along the edge results in a spin rotation effect under a magnetic field, which
is a manifestation of the Berry phase in the spin space. Besides, because the
external magnetic field lifts the spin degeneracy, a spin filter effect is
found, which arises from the spin-resolved resonant tunneling.

\section{The Hamiltonian and the Spin Rotation Mechanism}

The effective four-band Hamiltonian proposed for a HgTe/CdTe quantum well is
given by \cite{QSH}%
\begin{equation}
H(k)=\left(
\begin{array}
[c]{cc}%
h(k) & 0\\
0 & h^{\ast}(-k)
\end{array}
\right)  ,\label{QWH}%
\end{equation}
with $h(k)=\epsilon(k)I_{2\times2}+d_{\alpha}(k)\sigma^{\alpha},$
$\epsilon(k)=C-Dk^{2},$ $d_{\alpha}(k)=(Ak_{x},-Ak_{y},M(k)),$ and
$M(k)=M-Bk^{2}$, where we have used the basis order $\left\vert E_{1}%
+\right\rangle ,\left\vert H_{1}+\right\rangle ,\left\vert E_{1}-\right\rangle
,\left\vert H_{1}-\right\rangle .$ \textquotedblleft$E_{1}$\textquotedblright%
\ and \textquotedblleft$H_{1}$\textquotedblright\ represent the electron and
hole bands, \textquotedblleft$+$\textquotedblright\ and \textquotedblleft%
$-$\textquotedblright\ represent the spin along z direction \cite{JPSJ}. For a
strip geometry, it is well known the Hamiltonian with $M<0$ has a pair of
helical edge modes on each side of the strip. Similarly for a round disk
geometry we find it also has helical edge modes along the edge, each forming
an ideal 1D loop. The spin-up states travel clockwise, and the spin-down
states travel anti-clockwise.

When the Fermi energy of a QSH system is in the bulk gap, only the edge modes
are responsible to transport. The edge modes can be described by an effective
1D Hamiltonian\cite{Qi},%
\begin{equation}
H_{1D}(k)=v_{F}k\sigma_{z},\label{effective}%
\end{equation}
where $v_{F}$ is the Fermi velocity and $\sigma_{z}$ is the Pauli matrix
defined in the $(+,-)$ space. $k$ is the quasi-continuous momentum, whose
positive direction is defined along the azimuthal direction. In this 1D
Hamiltonian, $\sigma_{z}$ is a good quantum number, and the direction of the
motion is correlated to the spin polarization. Obviously this Hamiltonian
retains the information in the $(+,-)$ space and neglects the dynamics in the
($E_{1},H_{1}$) space, it can be used to study the properties of spin qualitatively.

Now we consider a QSH round connected to two normal electrodes without SOI, an
effect similar to the AB effect is expected. But the effect here must be
different from the conventional AB effect because the Hamiltonian
(\ref{effective}) is relative to spin. If the incident electron is polarized
in the x-y plane (we assume the quantum well is grown along the z direction),
the polarization direction of the outgoing current will rotate in the x-y
plane when an perpendicular magnetic field is applied on it. The magnetic flux
through the disk induces the spin current oscillation rather than the charge
current oscillation in the conventional AB effect. The spin rotation process
can be analyzed through the 1D Hamiltonian (\ref{effective}). An electron
injected from the left electrode polarized along positive x direction is
represented as a linear combination of spin-up and spin-down waves along the z
direction,
\begin{equation}
\left(
\begin{array}
[c]{c}%
1\\
1
\end{array}
\right)  =\left(
\begin{array}
[c]{c}%
1\\
0
\end{array}
\right)  +\left(
\begin{array}
[c]{c}%
0\\
1
\end{array}
\right)  ,
\end{equation}
When they enter the disk, they propagate in virtue of the edge states of the
QSH disk. The up-spin travels in the clockwise direction while the down-spin
travels in the anti-clockwise direction. When a homogeneous magnetic field
$\mathbf{B}=(0,0,\mathcal{B})$ is applied, the edge states can obtain
additional phases correlated with the vector potential $\mathbf{A}=\widehat
{e}_{\varphi}\mathcal{B}R_{eff}/2$ when they propagate along the edge, where
$\widehat{e}_{\varphi}$ is along the azimuthal direction, and $R_{eff}%
=\sqrt{\left\langle r^{2}\right\rangle }$ is the effective radius of an edge
state measured from the center of the disk to the ridge of the edge state. For
a large disk, $R_{eff}$ is about the radius of the disk $R$ because the edge
state's width is much smaller than $R$. The up- and down-spin obtain different
phases when they traveling in the opposite directions. At last they arrive at
the interface to the right electrode and recombine into a new wave,%
\begin{equation}
\left(
\begin{array}
[c]{c}%
e^{i(k+\frac{eA}{\hbar})\frac{L}{2}}\\
0
\end{array}
\right)  +\left(
\begin{array}
[c]{c}%
0\\
e^{i(k-\frac{eA}{\hbar})\frac{L}{2}}%
\end{array}
\right)  =e^{ik\frac{L}{2}}\left(
\begin{array}
[c]{c}%
e^{i\frac{eAL}{2\hbar}}\\
e^{-i\frac{eAL}{2\hbar}}%
\end{array}
\right)  ,
\end{equation}
where $L=2\pi R_{eff}$. The phase difference between up-spin and down-spin
determine the spin of the outgoing wave $\left\langle \overrightarrow
{\mathbf{\sigma}}\right\rangle =(2\cos(e\phi_{eff}/\hbar),-2\sin(e\phi
_{eff}/\hbar),0)$, where $\phi_{eff}=\mathcal{B}\pi R_{eff}^{2}$ is the
magnetic flux through the effective area enclosed by the edge states, so the
spin rotation angle is given by $\theta=\arctan(\left\langle \sigma
_{y}\right\rangle /\left\langle \sigma_{x}\right\rangle )=-e\phi_{eff}/\hbar$.
Taking into account the contribution from the dynamics in the ($E_{1},H_{1} $)
space which is omitted by the 1D Hamiltonian (\ref{effective}), the spin
rotation angle is given by%
\begin{equation}
\theta=\theta_{0}-2\pi\frac{\phi_{eff}}{\phi_{0}},\label{theta_analytic}%
\end{equation}
where $\phi_{0}=$ $h/e$ is the magnetic flux quantum, $\theta_{0}$ is the
intrinsic phase shift in the absence of the magnetic field which is related to
the details of the system such as the size, shape and interface. Yokoyama
\textit{et.al.} has studied the giant spin rotation happening at the interface
between a normal metal and a QSH insulator \cite{Nagaosa}. Here we focus on
the spin rotation induced by the magnetic field, so introduce $\theta_{0}$ as
a priori parameter.

\section{Numerical results}

\begin{figure}[ptb]
\begin{center}
\includegraphics[
trim=0.184354in 1.219219in 0.938691in 0.455986in, height=2.5071in,
width=3.1401in
]{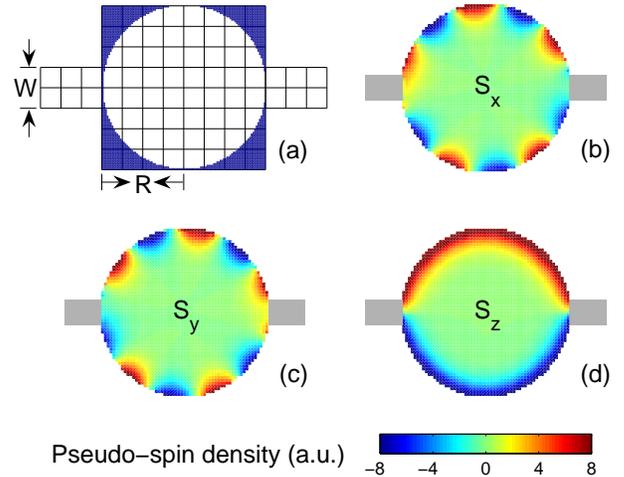}
\end{center}
\caption{(Color online) (a) Schematic diagram of the tight binding model for a
two-terminal QSH round disk device, the shadow area are assigned very high
on-site energy. The response of the pseudo-spin density $S_{x}$ (b), $S_{y}$
(c), $S_{z}$ (d) distribution in the disk, the gray regions represent the
leads.}%
\label{distribution}%
\end{figure}

We consider a two-terminal QSH round disk device connected with two normal
leads without SOI, as shown in Fig. \ref{distribution}(a). The disk is
described by the four-band Hamiltonian (\ref{QWH}), the leads are described by
a single-band Hamiltonian with spin degeneracy. The magnetic field
$\mathbf{B}=(0,0,\mathcal{B})$ is taken into account through the Peierls
substitution $\mathbf{k}\rightarrow\mathbf{k}+e\mathbf{A/\hbar}$ with the
Landau gauge $\mathbf{A}=(e\mathcal{B}y/\hbar,0,0)$, the Zeeman splitting is
ignored. In the tight-binding representation, the Hamiltonian of the whole
system is given by%
\begin{align}
H &  =H_{disk}+H_{lead}+H_{c},\label{Hamiltonian}\\
H_{disk}= &  \underset{\mathbf{i},\alpha}{\sum}V_{\alpha}c_{\mathbf{i}\alpha
}^{+}c_{\mathbf{i}\alpha}+\underset{\left\langle \mathbf{i},\mathbf{j}%
\right\rangle ,\alpha}{\sum}\frac{D_{\alpha}}{a^{2}}c_{\mathbf{i},\alpha}%
^{+}e^{i\mathbf{A\cdot(i-j)}}c_{\mathbf{j}\alpha}\nonumber\\
&  \mp\underset{\mathbf{i}}{\sum}A(A_{x}c_{\mathbf{i},E_{1}\pm}^{+}%
c_{\mathbf{i},H_{1}\pm}+h.c.)\nonumber\\
&  \mp\underset{\mathbf{i}}{\sum}A(\frac{1}{2ia}c_{\mathbf{i}+\delta_{x}%
,E_{1}\pm}^{+}c_{\mathbf{i},H_{1}\pm}+h.c.)\nonumber\\
&  +\underset{\mathbf{i}}{\sum}A(\frac{1}{2a}c_{\mathbf{i}+\delta_{y},E_{1}%
\pm}^{+}c_{\mathbf{i},H_{1}\pm}+h.c.),\\
H_{lead} &  =\underset{\mathbf{i},\beta}{\sum}V_{L}d_{\mathbf{i}\beta}%
^{+}d_{\mathbf{i}\beta}+\frac{D}{a^{2}}\underset{\left\langle \mathbf{i}%
,\mathbf{j}\right\rangle ,\beta}{\sum}d_{\mathbf{i}\beta}^{+}%
e^{i\mathbf{A\cdot(i-j)}}d_{\mathbf{j}\beta},\\
H_{c} &  =\frac{D}{a^{2}}\underset{\left\langle \mathbf{i},\mathbf{j}%
\right\rangle ,\alpha,\beta}{\sum}\left(  \gamma_{\alpha\beta}c_{\mathbf{i}%
\alpha}^{+}e^{i\mathbf{A\cdot(i-j)}}d_{\mathbf{j}\beta}+h.c.\right)  ,
\end{align}
where $H_{disk},$ $H_{lead},$ $H_{c}$ are for the disk, leads and the
disk-lead coupling, respectively. $\alpha\in\{E_{1}\pm,H_{1}\pm\}$ and
$\beta\in\{\uparrow,\downarrow\}$. $c_{\mathbf{i}\alpha}$ annihilates an
$\alpha$-band eletron on the site $\mathbf{i}=(i_{x},i_{y}\mathbf{)}$ in the
disk, and $d_{\mathbf{i}\beta}$ annihilates a spin-$\beta$ electron on the
site $\mathbf{i}$ in the leads. $a=3.3nm$ is the lattice constant, and
$\delta_{x(y)}$ is the unit vector along the x(y) direction. $D_{\alpha}=D+B$
for $\alpha\in\{E_{1}\pm\}$ and $D_{\alpha}=D-B$ for $\alpha\in\{H_{1}\pm\}$.
$\gamma_{\alpha\beta}$ describes the coupling between the $\alpha$-band in the
disk and the spin-$\beta$ in the leads. If the interfacial spin-flip
scattering is neglected, it can be estimated $\left\{  \gamma_{E_{1}%
+,\uparrow},\gamma_{H_{1}+,\uparrow},\gamma_{E_{1}-,\downarrow},\gamma
_{H_{1}-,\downarrow}\right\}  \gg\left\{  \gamma_{E_{1}+,\downarrow}%
,\gamma_{H_{1}+,\downarrow},\gamma_{E_{1}-,\uparrow},\gamma_{H_{1}-,\uparrow
}\right\}  $ by comparing the spin expectation values of the bases
\cite{QSH,8band}. In the following calculation, we assume $\gamma
_{E_{1}+,\uparrow}=\gamma_{H_{1}+,\uparrow}=\gamma_{E_{1}-,\downarrow}%
=\gamma_{H_{1}-,\downarrow}=\gamma$ and $\gamma_{E_{1}+,\downarrow}%
=\gamma_{H_{1}+,\downarrow}=\gamma_{E_{1}-,\uparrow}=\gamma_{H_{1}-,\uparrow
}=0$. The on-site energy in the disk is given by
\begin{equation}
V_{\alpha}=\left\{
\begin{array}
[c]{c}%
C+M-\frac{2(D+B)}{a^{2}},\alpha\in\{E_{1}\pm\}\\
C-M-\frac{2(D-B)}{a^{2}},\alpha\in\{H_{1}\pm\}\\
+\infty,\mathbf{i}\in\{\text{shadow area}\}
\end{array}
\right.  ,
\end{equation}
which promises the effective geometry is a round disk. The on-site energy in
the leads is $V_{L}=C_{L}-\frac{4D}{a^{2}},$ and $C_{L}=-0.1eV$ promises Fermi
energy of the leads is in the conduction band \cite{Nagaosa}. The other
parameters we used are $A=364.5meV\cdot nm,B=-686meV\cdot nm^{2}%
,C=0,D=-512meV\cdot nm^{2},M=-10meV$ \cite{JPSJ}. The radius of the disk is
$R=100nm$, and the width of the leads are $L=30nm$.

We assume the left lead has a x-spin-polarized potential V, the
x-spin-resolved Fermi energy in the leads is $E_{f}^{L\rightarrow}%
-V=E_{f}^{L\leftarrow}=E_{f}^{R\rightleftarrows}\equiv E_{f}$. The spin
conductance with respect to the spin along $\alpha$-axis in the right lead is
defined by $G^{\alpha}=\lim_{V\rightarrow0}J^{\alpha}/V$, where $J^{\alpha}$
is the $\alpha$-spin current in the right lead. The practical calculation
bases on the lesser Green's function in the right lead,%
\begin{equation}
G^{\alpha}(i_{x})=\frac{2e^{2}}{h}\underset{i_{y}}{\sum}\operatorname{Re}%
\left[  Tr(\sigma_{\alpha}H_{\mathbf{i}+\delta_{x},\mathbf{i}}G_{\mathbf{i}%
,\mathbf{i}+\delta_{x}}^{<})\right]  ,\label{Gax}%
\end{equation}
where the lesser Green's function $G^{<}=G^{r}\Sigma_{L\rightarrow}^{<}G^{a},
$ in which $G^{r,a}=\left[  E-H\pm i0^{+}\right]  ^{-1}$, and the lesser
self-energy $\Sigma_{\rightarrow}^{<}=\frac{\gamma^{2}D^{2}}{a^{4}}%
U^{+}(G_{L\uparrow}^{a}-G_{L\uparrow}^{r})U$. $G_{L\uparrow}^{a}=\left(
G_{L\uparrow}^{r}\right)  ^{+}$ and $G_{L\uparrow}^{r}$ is the surface Green's
function in the z-spin-up subspace for a semi-infinite lead obtained by a
recursive method \cite{lead}. $U=e^{i\sigma_{y}\pi/4}$ is a $SU(2)$ rotation
from z-spin to x-spin. Although Eq. (\ref{Gax}) is a function of the
x-coordinate, the result is independent of $i_{x}$ because the spin
conversation in the normal leads without SOI.

\begin{figure}[ptb]
\begin{center}
\includegraphics[
trim=0.051395in 0.000000in 0.000000in 0.026173in, height=2.2269in,
width=3.1401in
]{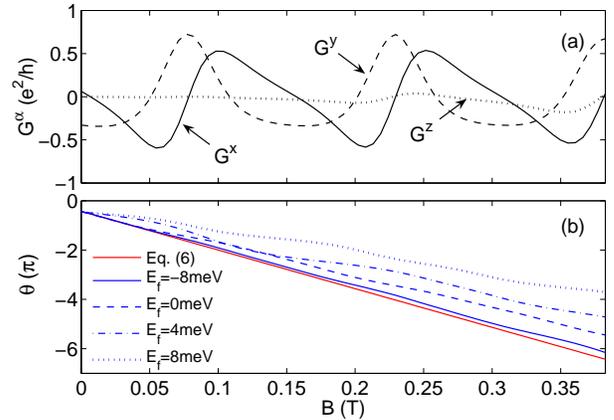}
\end{center}
\caption{(Color online) (a) Spin conductances $G^{x}$ (solid line), $G^{y}$
(dashed line), $G^{z}$ (dotted line) as functions of the magnetic field for
$E_{f}=-0.8meV$. (b) The rotation angle as a function of the magnetic field
for various Fermi energy, the red solid line is from the analytical formula
Eq. (\ref{theta_analytic}) with $\theta=-0.43\pi$ and $R_{eff}=R$. The blue
solid, dashed, dot-dashed, dotted lines are from the numerical calculations
for $E_{f}=-8,0,4,8meV,$ respectively.}%
\label{rotation}%
\end{figure}

Fig. \ref{rotation}(a) shows the spin conductances as functions of the
magnetic field for a fixed Fermi energy $E_{f}=-0.8meV.$ The oscillations on
$G^{x},G^{y}$ means a spin rotation in the x-y plane. The spin rotation angle
is defined by $\theta=\arctan(G^{y}/G^{x})$. Fig. \ref{rotation}(b) shows the
rotation angles as functions of the magnetic field for various Fermi energy
$E_{f}=-8,0,4,8meV$. As a comparison, the red solid line shows the analytical
result from Eq. (\ref{theta_analytic}) with the value assignments $\theta
_{0}=-0.43\pi$ and $R_{eff}=R$. The numerical curves follow the same tendency
with the analytical line. The fluctuations on the numerical curves are caused
by the evolution in the ($E_{1},H_{1}$) space which is not included in Eq.
(\ref{theta_analytic}). The deviations between the numerical curves and the
analytical line are because the effective radius of the edge states is smaller
than the disk radius. We use Eq. (\ref{theta_analytic}) fitting the numerical
curves to get $R_{eff}/R=0.98,0.91,0.85,0.75$ for $E_{f}=-8,0,4,8meV$. In the
energy regime we study, a smaller effective radius is deduced for a higher
Fermi energy, which is consistent with the results from Chu \textit{et.al.}
\cite{Churuilin}. The edge states with higher energy are wider and penetrate
more towards the center of the disk, so the enclosed area have a smaller
effective radius $R_{eff}$. The phenomenon of the spin rotation angle
depending on the Fermi energy is called the \textit{dispersibility }of the
spin rotation effect, on the contrary to the nondispersibility of the
conventional AB effect in which the conductance oscillation is independent of
the energy of the particles \cite{nondispersive}. The dispersibility here is
due to the effective 1D ring formed by the edge state has a energy-dependent area.

A similar spin rotation has been reported in the scalar AB experiment with
Neutrons\cite{Allman}, but the result is explained unnecessarily by the phase
shift difference of the AB effect. It can be explained by a classical spin
procession instead \cite{comment} We emphasize the spin rotation effect we
discover here is a topological effect which can not be explained by the
classical spin precession. It is a manifestation of Berry's topological phase
in the spin space, due to the phase interference of the different z-spin
components, bearing an analogy to the rotation of linearly polarized light in
a helically wound optical fiber \cite{Wuyongshi,Chiao}.

To see the transport process in the disk detailedly, we calculate the
pseudo-spin density distribution $\mathbf{S(i)}$ in Fig. \ref{distribution},
as a response of the x-spin-polarized potential in the left lead. The
pseudo-spin is defined in the disk region as $\widehat{S}_{\alpha}%
=I_{2}\otimes\sigma_{\alpha}$, where $I_{2}$ is the unit matrix acting on the
($E_{1},H_{1}$) space and $\sigma_{\alpha}$ is the Pauli matrix acting on the
$(+,-)$ space. This definition is just the \textquotedblleft
spin\textquotedblright\ defined by Yokoyama \textit{et.al.} \cite{Nagaosa}. We
note here that the real spin is not a good quantum number in the QSH insulator
because of the spin-orbit coupling. So the pseudo-spin is used to distinguish
the two states of a Kramers doublet. We observe the pseudo-spin density
distribution in the disk to learn about the way of the electrons' propagation.
We choose $B=0.0785T$ and $E_{f}=-8meV$, where $G^{y}=0.75e^{2}/h$,
$G^{x}\approx G^{z}\approx0$ in Fig. \ref{rotation}(a). The pseudo-spin
density response is calculated through the lesser Green's function,%
\begin{equation}
S_{\alpha}(\mathbf{i})=-iTr\left[  \widehat{S}_{\alpha}G_{\mathbf{ii}}%
^{<}\right]  .\text{ }(\alpha=x,y,z)
\end{equation}
We can see that the pseudo-spin density concentrates on the edge of the disk,
which ensures the transport is born by the edge states. Although the incident
electrons are polarized along x direction, the local spin on the edge is
mainly polarized in z direction, and the upper edge and the lower edge are
occupied by the opposite polarized z-spins. The edge states with opposite
z-spins acquire different phase shifts when they travel along the opposite
edges, resulting in an overall spin rotation in the x-y plane after
superposition at the interface to the right lead. So it is the manifestation
of AB effect in the spin space. Besides, we can observe the x- and y-spin wave
along the edge, which is also arising from the interference of the z-spin-up
and -down wavefunctions because a part of the electron wave is reflected at
the interfaces to the leads, forming a standing wave along the edge and
interfering with each other.

\section{Spin filter effect}

The QSH round disk has a spin filter effect for the spin in the z direction
when a magnetic field lifts the spin degeneracy. The mechanism for lifting the
spin degeneracy is not the tiny Zeeman splitting but the angular momentum of
the edge state. The edge state travels closely to the boundary of the disk, so
it has a large angular momentum which strongly couples to the extermal
magnetic field. We can estimate the energy levels for the edge states of a QSH
round disk. Zhou \textit{et.al.} has solved the dispersions of the edge states
of a QSH strip exactly\cite{Zhoubin},%
\begin{equation}
E_{\pm}(k)=-\frac{MD}{B}\pm A\sqrt{\frac{B^{2}-D^{2}}{B^{2}}}k,
\end{equation}
where $E_{+}(E_{-})$ derives from the upper (lower) block of the Hamiltonian
(\ref{QWH}). For a large round disk, an analytical solution gives the
quantization condition $k\approx(n+1/2)/R_{eff}$ \cite{Shanwenyu}. When a
magnetic field $\mathbf{B}=(0,0,\mathcal{B})$ is applied, the energy level
shift in the linear order of $\mathbf{B}$ is given by%
\begin{equation}
\Delta E_{\pm}=\frac{e}{2}\left\langle \mathbf{r\times v}\right\rangle
\mathbf{\cdot B}\approx\mp\frac{e\mathcal{B}}{2\hbar}\frac{\partial
E}{\partial k}R_{eff}.
\end{equation}
So the energy dispersions for the edge states are approximately given by%
\begin{equation}
E_{\pm}(n)=-\frac{MD}{B}+A\sqrt{\frac{B^{2}-D^{2}}{B^{2}}}\left(  \frac
{2n+1}{2R_{eff}}\mp\frac{e\mathcal{B}}{2\hbar}R_{eff}\right)  .\label{En}%
\end{equation}

\begin{figure}[ptb]
\begin{center}
\includegraphics[
trim=0.000000in 0.000000in 0.350624in 0.163170in, height=2.4491in,
width=3.1721in
]{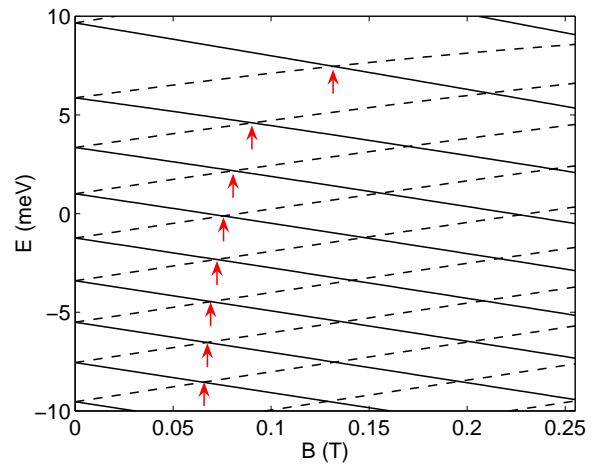}
\end{center}
\caption{(Color online) Energy levels as functions of the magnetic field. The
solid lines are from the upper block $h(k)$ and the dashed lines are from the
lower block $h^{\ast}(-k)$. The red arrow indicate a series of crossing
points.}%
\label{level}%
\end{figure}

For a disk with $R=100nm,$ the dispersions calculated by the tight-binding
method are shown in Fig. \ref{level}, which are in agreement with Eq.
(\ref{En}) very well. The solid lines are from the upper block $h(k)$ and the
dashed lines are from the lower block $h^{\ast}(-k)$ in Hamiltonian
(\ref{QWH}). The energy levels are linear functions of the magnetic field, the
spin-up levels and spin-down levels shift oppositely and cross each other. The
red arrows indicate a series of crossing points. From the energy dispersions
Eq. (\ref{En}), we can estimate the crossing points locate at $\phi
_{eff}(n)=n\phi_{0}/2$. In Fig. \ref{level}, the crossing points shift right
when the energy is higher, that means the effective radius of a higher-energy
edge state is smaller, that is consistent with the deductions from Fig.
\ref{rotation}(b) and Ref. \onlinecite{Churuilin}.

\begin{figure}[ptb]
\begin{center}
\includegraphics[
trim=0.000000in 0.370642in 0.372209in 0.187295in, height=2.2096in,
width=3.1401in ]{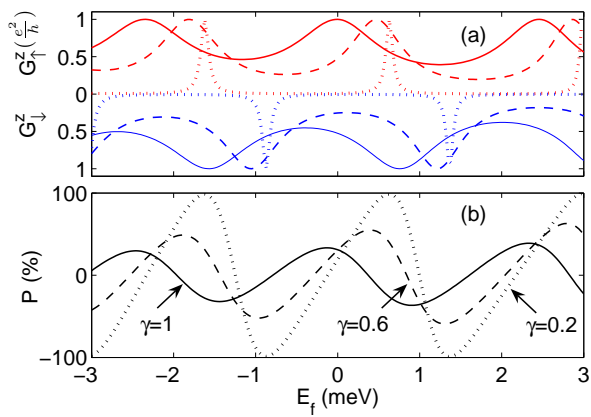}
\end{center}
\caption{(Color online) (a) The z-spin resolved conductance as a function of
the Fermi energy with a magnetic field $\mathcal{B}=0.025T$, the red (blue)
lines represent $G_{\uparrow}^{z}$ $(G_{\downarrow}^{z}).$ (b) The z-spin
polarization ratio of the conductance as a function of the Fermi energy. In
the both subgraphs, the solid, dashed, dash-dotted lines are for the disk-lead
coupling $\gamma=1,0.6,0.2,$ respectively.}%
\label{Gz}%
\end{figure}

The lifting of the z-spin degeneracy in a magnetic field can induce a spin
filter effect. If we consider the round disk as a quantum dot, the tunneling
current through it must be z-spin polarized. We use the non-equilibrium
Green's function to calculate the z-spin resolved conductance. Still for the
geometry in Fig. \ref{distribution}(a), the left lead has a higher Fermi
energy than the right lead, $E_{f}^{L}-V=E_{f}^{R}\equiv E_{f}$, and the Fermi
energy is spin-independent so that the incoming current is unpolarized. The
z-spin-resolved conductance is defined by $G_{\alpha}^{z}=\lim_{V\rightarrow
0}J_{\alpha}^{z}/V$, where $J_{\uparrow}^{z}(J_{\downarrow}^{z})$ is the
z-spin-up(down) component of the outgoing current in the right lead. The
practical calculation bases on the lesser Green's function,%
\begin{align}
G_{\alpha}^{z}(i_{x})  &  =\frac{2e^{2}}{h}\underset{i_{y}}{\sum
}\operatorname{Re}\left[  Tr(\sigma_{\alpha}^{z}H_{\mathbf{i}+\delta
_{x},\mathbf{i}}G_{\mathbf{i},\mathbf{i}+\delta_{x}}^{<})\right]  ,\\
\sigma_{\uparrow}^{z}  &  =\left(
\begin{array}
[c]{cc}%
1 & 0\\
0 & 0
\end{array}
\right)  ,\sigma_{\downarrow}^{z}=\left(
\begin{array}
[c]{cc}%
0 & 0\\
0 & 1
\end{array}
\right)  .
\end{align}
Same to Eq. (\ref{Gax}), the results are independent of $i_{x}$ because of the
spin conservation in the leads. The charge conductance $G_{c}=G_{\uparrow}%
^{z}+G_{\downarrow}^{z}$, and the z-spin conductance $G^{z}=G_{\uparrow}%
^{z}-G_{\downarrow}^{z}$. The z-spin-resolved conductances as functions of the
Fermi energy are shown in Fig. \ref{Gz}(a) with three different disk-lead
coupling $\gamma$ in the Hamiltonian (\ref{Hamiltonian}). The magnetic field
is fixed at $\mathcal{B}=0.025T$, and the Fermi energy can be adjusted by a
gate voltage in experiments. The spin-resolved resonant tunneling happens when
the Fermi energy is aligned with any energy level of the disk. We define the
spin polarization ratio of the conductance $P=G^{z}/I $, which is shown in
Fig. \ref{Gz}(b) for the corresponding $\gamma$. For a smaller $\gamma$, the
conductance peaks are narrower, so the separation of the spin-up and spin-down
conductances is clearer, thus the spin polarization ratio is larger. At some
Fermi energy, the spin polarization ratio reaches $\pm100\%$ when $\gamma
=0.2$. At this time, only one edge state with a specific z-spin takes charge
of the transport, so the outgoing current in the right lead is totally
polarized, the QSH round disk exhibits a perfect spin filter effect.

\bigskip

\section{Conclusion}

In conclusion, we study the spin transport of a QSH round disk device
connected to two normal electrodes without SOI. An in-plane spin rotation
effect is discovered as a manifestation of the AB effect in the spin space.
This phenomenon arises from the interference of the z-spin-up and -down edge
states which acquire different phase shifts when traveling in the opposite
directions. The spin rotation angle is related to the magnetic flux through
the effective area enclosed by the helical edge states.

Besides the spin rotation effect, the QSH round disk has a z-spin filter
effect when the bias voltage is spin-independent. The effect is due to the
electron's tunneling through the disk via its discrete spin-separating energy
levels when a magnetic field lifts the spin degeneracy of the system. This
effect is notable when the couplings between the disk and leads are weak. For
a small disk-lead coupling, the z-spin polarization ratio of the outgoing
current can reach $\pm100\%$ if an appropriate gate voltage and magnetic field
are applied.

The two effects are both induced by the helical edge states of the QSH
material, they are not sensitive to the size and the shape of the disk as long
as the circumference of the disk is smaller than the coherent length of the
edge states. We hope the effects discovered here can help designing
multi-functional spintronic devices and show technological applications in the
quantum computing and microelectronics.

\acknowledgments{The authors thank S. Q. Shen, H. Z. Lu, R. L. Chu
and J. Li for helpful discussions. This work was supported by the
Research Grant Council of Hong Kong under Grant No.: HKU 7041/07P
and HKU 10/CRF/08.}

\bigskip

\textit{Note added}.--After the completion of this work, we became aware of a
recent preprint by J. Maciejko \textit{et. al.}\cite{Qixiaoliang}, which
proposes a spin transistor based on the spin rotation effect (spin AB effect).

\end{document}